\documentclass[preprint,prl,aps,superscriptaddress,showpacs]{revtex4}

\usepackage{amsmath}
\usepackage{graphicx}

\begin{document}

\title{Which is the temperature of granular systems?
A mean field model of free cooling inelastic mixtures.}

\author{Umberto Marini Bettolo Marconi}
\affiliation{
Dipartimento di Matematica e Fisica \\
Universit\'a di Camerino, 62032 Camerino, Italy}
\affiliation{
Istituto Nazionale di Fisica della Materia, Unit\`a di Camerino, Camerino, Italy}

\author{Andrea Puglisi} 
\affiliation{
Dipartimento di Fisica \\
Universit\'a ``La Sapienza'',P.le A. Moro 2, 00198 Roma, Italy}
\affiliation{
Istituto Nazionale di Fisica della Materia, Unit\`a di Roma, Roma, Italy}

\date{\today}

\begin{abstract}
We consider a mean field model describing the free cooling process
of a two component granular mixture, a generalization of so
called Maxwell model.
The cooling is viewed as an ordering process and the scaling
behavior is attributed to the presence of an attractive
fixed point at $v=0$ for the dynamics. 
By means of asymptotic
analysis of the Boltzmann equation and of numerical simulations
we get the following results:

1)we establish the existence of two different  partial  granular temperatures,
one for each component, which violates the Zeroth Law of Thermodynamics;
2) we obtain the scaling form of the two distribution functions;
3) we prove the existence of a continuous spectrum of exponents characterizing
the inverse-power law decay of the tails
of the velocity , which generalizes the
previously reported value $4$ for  the pure model;
4) we find that 
the exponents depend on the composition, masses and restitution coefficients
of the mixture;
5) we also remark 
that the reported distributions represent a dynamical realization
of those predicted by the Non Extensive Statistical Mechanics,
in spite of the fact that ours 
stem from a purely dynamical approach.
\end{abstract}

       
\maketitle


\section{Introduction}

One of the basic principles over which the Thermodynamics is built,
derives from the simple observation that when two systems A and B are
brought together so that they can exchange energy (or in thermal
contact), after some time they reach a mutual macroscopic equilibrium
state in the absence of energy sources, i.e. their temperatures become
equal. A second observation is that the thermal equilibrium between a
third system C and A implies the thermal equilibrium between B and
C. These two facts, which represent the content of the Zeroth Law of
Thermodynamics, give sense to the concept of temperature and allow us
to build thermometers \cite{Pippard}. The simplest state of
aggregation of matter, namely the gaseous state, offers a neat example
of the validity of such a principle. When in equilibrium the state of
a mixture of different gases is characterized by a single temperature,
i.e. every species has the same average kinetic energy per particle
regardless its molecular nature.

An interesting question related to the Zeroth Law arises quite
naturally when we consider the behavior of Granular Gases, as
assemblies of moving grains at low density are usually called. More in
general, Granular Matter represents a relatively new and unexplored
area of research which has attracted the attention of physicists. As
ordinary matter Granular Matter may appear under different guises:
solid-like or dense aggregates of particles where the motion of its
constituents is negligible, liquid-like when a dense aggregate flows,
and as a gas when the mutual distances between grains are larger that
their typical linear size. Of course, such a classification is
empirical but quite loose, since we are not allowed to employ concepts
such as thermodynamic phases, in fact Granular Matter does not fall into
the realm of Thermodynamics. In other words, only if we are able to
show that the basic Laws of Thermodynamics work in the case of
granular materials we can employ its results. There is an emerging
consensus that granular materials do not achieve a proper thermodynamic
equilibrium. Hence, even if the temperature of a
granular assembly can be suitably defined, it does not have the
properties required by standard thermodynamics, as we shall discuss
below.  We shall not digress on some interesting recent proposals
aimed to define the temperature of granular packings \cite{Edwards},
which is a measure of the potential energy landscape, but we shall
confine our attention to dilute systems of granular particles, 
where the term
granular temperature, defined exactly in the same way as in ordinary
gases, i.e.  as the average of the kinetic energy per particle, has
been coined~\cite{ogawa}.
How far pursue such an analogy?. Is the granular
temperature a common feature of all granular gases in mutual
equilibrium, i.e. the quantity which has the role of determining if
two system are in equilibrium with respect to each other with respect
to exchanges of energy?  If it is not so, the granular temperature has
to be regarded just as an ensemble average of the kinetic energy of
the grains, which may depend on the microscopic details and therefore
is spoiled of one of its most useful characteristics.

In the present paper we shall deal with such an issue through a simple
model of granular mixture. To be more specific we shall describe a
granular mixture by means of a model of the Maxwell type, i.e. a model
characterized by a collision rate which is independent of the relative
velocity of the colliding particles \cite{Ernst1}.  Such a class of
models has its justification in the case of elastic Maxwell molecules,
i.e. particles interacting via a soft repulsive pair potential of the
form $r^{1-2 d}$. In the case of inelastic particles the constant
collision rate is just a matter of mathematical convenience, since it
does not correspond to any known microscopic interaction.  In spite of
that, it has been demonstrated by several authors
\cite{Bobylev},\cite{Cercignani}\cite{BenNaim},\cite{Balda2}
\cite{Ernst2}, that the cost of
sacrificing physical realism has been widely compensated by the amount
of exact analytical results and simplification even in the numerical
treatment of the collision process.  On the other hand the real
question is: what are the physical consequences of the constant
collision rate on our predictions?  Some of these are robust with
respect to the choice of the model others will be peculiar, as we shall
discuss below. An interesting
observation is that the homogeneous cooling states
of Maxwell models and of more
realistic models become quite similar if one measures
the respective evolutions in terms of the number of collisions 
occurred up to a given
instant, while they are very different in the presence
of spatial inhomogeneities \cite{Balda2},
\cite{Vannoije},\cite{Balda1},\cite{Umbm}.
In the case of granular mixtures there is another aspect
which is robust with respect to the
choice of the model, namely the existence of a two temperature behavior. 
In
fact, this has already been observed by Dufty and Garz\'o \cite{Dufty} in a
study of the cooling of a mixture based on the Enskog approximation
and experimentally by Feitosa and Menon \cite{Menon} in a mixture
subject to external driving. The Maxwell model, we shall study, also
shows such a behavior, but offers the advantage that, being
considerably simpler than other approaches, provides
a guidance to understand the global aspect of the problem.

Summarizing, the choice of the model is deliberately minimalistic for
the following three important reasons:

a) the simplicity of the model allows us to avoid the usual problem of
the closure of the correlation hierarchy in statistical physics and
the need of introducing approximations. In fact, the Boltzmann
equation turns out to be exact for the model we study, so that none of
the results we find can be ascribed to a breakdown 
under some peculiar conditions
of some physically
motivated approximation, such as the Boltzmann molecular chaos
hypothesis;

b) most of the work can be performed analytically and even the form of
the asymptotic solutions of the Boltzmann equation can be estimated in
closed manner with a fairly good accuracy;

c) our solutions can be validated by simple simulations of the
microscopic collision process.

Perhaps it is worth to mention that as an unexpected byproduct of our
study we have found a rich variety of behaviors of the velocity
distributions as a function of the microscopic control parameters. The
common feature of all these distributions is the presence of inverse
power law high velocity tails. These arise quite naturally from our
microscopic dynamics and do not need any fine tuning of the
parameters, a feature somehow similar to 
that which characterize the so called  Self Organized Criticality.

Incidentally, the observed inverse-power laws for the distributions
are identical to those predicted by the Non Extensive Statistical
Mechanics (NESM) approach \cite{Tsallis} \cite{Tsallis2}, which apart
from some exceptions still lacks of dynamical foundations.  An
important point to stress is the fact that the observed behavior
derives from the solution of the Boltzmann transport equation for a
system clearly inspired to the physics of granular materials and
already employed in the past, and not the result of assumptions about
the form of the generalized entropy, as we shall see below.
  
The structure of the present paper is the following: in section II we
introduce the model, define the notation and write the Boltzmann equation;
in section III we set up the moment expansion and characterize by a
granular temperature the macroscopic state of each component; sections
IV and V, in which we discuss how the scaling solutions for the
Boltzmann equation can be obtained, contain all technical details; in
section VI we comment the connections between our results and the the
NESM approach; in VII we present our conclusions.
 

\section{Definition of the model}

We shall consider an assembly of $N_1$ particles of species $1$ and
$N_2$ particles of species $2$ endowed with scalar velocities
$v_i^{\alpha}$, with $\alpha=1,2$. The two species may have different
masses, $m_1$ and $m_2$ and/or different restitution coefficients
$r_{11}, r_{22}, r_{12}=r_{21}$.

The main assumptions of the model are:

a) the forces are pairwise and impulsive, so that 
their velocities change at each collision according to the
following rule:

\begin{subequations}
\label{collision}
\begin{align}
v_i^{'\alpha} & =  v_i^{\alpha}- [1+r_{\alpha \beta}]\frac{m_{\beta}}{m_{\alpha}+ m_{\beta}}(v^{\alpha}_i-v^{\beta}_j) \\
v^{'\beta}_j & =  v^{\beta}_j+ [1+r_{\alpha \beta}]\frac{m_{\alpha}}{m_{\alpha}+m_{\beta}}(v^{\alpha}_i-v^{\beta}_j)
\end{align}
\end{subequations}

b) collisions involving simultaneously more than two particles are
disregarded;

c) within the spirit of mean field models all pairs are allowed
to exchange momentum regardless of their mutual positions.

In eqs. \eqref{collision}
the primed quantities are the post-collisional velocities and are a
linear combination of the velocities before the collisions. The
process conserves the number of particles of each species and the
total impulse $\Pi=\sum_{\alpha=1,2} \sum_{i=1}^{N_{\alpha}}
m_{\alpha} v^{\alpha}_i$.

To study the statistical behavior of the system we consider the
coupled equations for the velocity probability distribution functions
$P_{\alpha}(v,t)$, which give the probability density of finding a
particle of species $\alpha$ with velocity $v$ at the instant $t$. In
the absence of driving forces we write the following two component
Boltzmann equation:

\begin{equation}
\partial_t P_i(v_1,t)\!=\! \sum_{j=1}^{2} \int dv_2
[\frac{1}{\alpha}P_i(v_1'')P_j(v_2'') - P_i(v_1)P_j(v_2)].
\label{boltzmann}
\end{equation}

The system of Boltzmann equations \eqref{boltzmann} describes the
evolution of the p.d.f. of the velocities in the case of $N_1$ and
$N_2$ going to infinity. From a rigorous point of view correlations
can appear in a finite system even if such a mean field approach is
employed, i.e.  the Molecular Chaos assumption underlying the Boltzmann
equation is no more guaranteed to be true for finite $N_1$ and
$N_2$. The typical mechanism of failure of the Molecular Chaos is the
re-collision process: the collisional history of two colliding
particles may overlap and therefore the two colliding particles are
somehow correlated, e.g.  the two particles may have collided together
in the recent past or may have collided with different particles that
have collided together at previous times (ring collisions). In the
simulations we have always used at least one million of particles: this
is enough, we think, to expect a good comparison with analytical
results.

Eliminating the pre-collisional velocities $v_1'',v_2''$ in favor of
the post-collisional velocities by means of eq. \eqref{collision} we
obtain the non linear system:

\begin{subequations}
\label{prob}
\begin{gather}
\begin{split}
\partial_t P_1(v,t)\!&=-\!P_1(v,t)\!+\!\frac{2 p}{1+r_{11}}
\!\int\!\!\!du\, 
P_1(u,t)P_1\left(\frac{2 v-(1-r_{11})u}{1+r_{11}},t\right) \\
&+\frac{(1-p)}{1+r_{12}}\frac{m_1+m_2}{m_2} \int\!\!\!du\,
P_1(u,t)P_2\left(\frac{\frac{m_1+m_2}{m_2}
v-(\frac{m_1}{m_2}-r_{12})u}{1+r_{12}},t\right)
\end{split} \\
\begin{split}
\partial_t P_2(v,t)\!&=-\!P_2(v,t)\!+\!\frac{2 (1- p)}{1+r_{22}}
\!\int\!\!\!du\,P_2(u,t)P_2\left(\frac{2 v-(1-r_{22})u}{1+r_{22}},t\right) \\
&+\frac{p}{1+r_{12}}\frac{m_1+m_2}{m_1} \int\!\!\!du\,
P_2(u,t)P_1\left(\frac{\frac{m_1+m_2}{m_1}
v-(\frac{m_2}{m_1}-r_{12})u}{1+r_{12}},t\right).
\end{split}
\end{gather}
\end{subequations}

The generalization to many components is straightforward.

To proceed further, it is useful to employ the method of
characteristic functions \cite{Bobylev},\cite{BenNaim}, i.e. the Fourier
transforms of the probability densities defined as:

\begin{equation}
\hat P_{\alpha}(k,t)=\int_{-\infty}^{\infty} dv e^{ikv} P_{\alpha}(v,t)
\label{character}
\end{equation}

The integral equations in Fourier space assume the more convenient
form :

\begin{subequations}
\label{fou}
\begin{gather}
\begin{split}
\partial_t \hat P_1(k,t)&=-\hat P_1(k,t)+ p \hat P_1(\gamma_{11}k,t)  \hat P_1((1-\gamma_{11})k,t) \\
&+(1-p) \hat P_1(\tilde\gamma_{12}k,t) \hat P_2(1-\tilde\gamma_{12})k,t) 
\end{split} \\
\begin{split} \label{fou2}
\partial_t \hat P_2(k,t)&=-\hat P_2(k,t)+ (1-p) \hat P_2(\gamma_{22}k,t)  \hat P_2((1-\gamma_{22})k,t) \\
&+p \hat P_2(\tilde\gamma_{21}k,t)\hat P_1((1-\tilde\gamma_{21})k,t) 
\end{split}
\end{gather}
\end{subequations}

where $p=N_1/(N_1+N_2)$, $\zeta=m_1/m_2$ and 

\begin{subequations}
\begin{align}
\gamma_{\alpha \beta} &=\frac{1-r_{\alpha \beta}}{2} \\
\tilde\gamma_{12} &=[1-\frac{2}{1+\zeta}(1-\gamma_{12})] \\
\tilde\gamma_{21} &=[1-\frac{2}{1+\zeta^{-1}}(1-\gamma_{12})] 
\end{align}
\end{subequations}

Interestingly, a mixture of elastic particles of unequal masses does
equilibrate under the evolution rule \eqref{collision}, unlike the
case $\zeta=1$. In fact, the solutions of \eqref{fou} are Gaussians
$e^{-\frac{\alpha}{m_{\alpha}} k^2}$, and in real space correspond to
Maxwellian velocity distributions.


\section{The moment expansion of the distribution functions}

For finite inelasticity, let us tentatively assume $\hat
P_{\alpha}(k,t)$ to be analytic at the origin $k=0$ and perform a
Taylor expansion in powers of $k$:

\begin{equation}
\hat P_{\alpha}(k,t)=\sum_{n=0}^{\infty} \frac{(i k)^n}{n !} \mu^{\alpha}_{n}(t)
\label{Taylor}
\end{equation} 

The coefficient $\mu^{\alpha}_{n}(t)$ of order $n$ of the expansion
represents the n-th moment of the distribution $P_{\alpha}(v,t)$
according to the definition eq. \eqref{character}.  By substituting
the Taylor expansion and equating coefficients of powers of $k$ to
zero one obtains a set of linear equations for the moments which can
be systematically solved.  The moments of order zero are
$\mu^{\alpha}_0(t)=1$, due to the normalization of $P_{\alpha}(v,t)$ ,
while those of first order are solutions of the coupled linear
equations:

\begin{subequations}
\label{prim}
\begin{align}
\partial_t \mu^{1}_{1}(t)&= -(1-p)(1-\tilde\gamma_{12}) [\mu^{(1)}_{1}-\mu^{(2)}_{1}] \\
\partial_t \mu^{2}_{1}(t)&= p(1-\tilde\gamma_{21}) [\mu^{(1)}_{1}-\mu^{(2)}_{1}]
\end{align}
\end{subequations}

The total impulse $\Pi=N[p m_1 \mu^{(1)}_{1}+(1-p)m_2 \mu^{(2)}_{1}]$
is a constant of motion and corresponds to the eigenvalue $z_1=0$
associated with the Galilean invariance, of the linear system
\eqref{prim}, whereas the negative eigenvalue
$z_2=-(1+r_{12})[\frac{m_1 p+(1-p) m_2}{m_1+m_2}]$, describes the
exponential law $e^{-z_2 t}$ by which two subsystems $1$ and $2$ in
relative motion with respect to each other come to have the same
average velocity.
 
The second order cumulants represent the most important statistical
indicator of the state of a granular gas and are usually taken as the
definition of partial granular temperatures . In order to study their
evolution one has to solve the following coupled linear equations for
the second moments:

\begin{subequations}
\label{secon}
\begin{align}
\partial_t \mu^{(1)}_{2}(t)&=d^{(2)}_{11} \mu^{(1)}_{2}+ d^{(2)}_{12}  \mu^{(2)}_{2} \\
\partial_t \mu^{(2)}_{2}(t)&=d^{(2)}_{21} \mu^{(1)}_{2}+ d^{(2)}_{22} \mu^{(2)}_{2}
\end{align}
\end{subequations}

where the coefficients are given by:

\begin{subequations}
\begin{align}
d^{(n)}_{11}&=-1+p[\gamma_{11}^n+(1-\gamma_{11})^n] +(1-p)[\tilde\gamma_{12}]^n \\
d^{(n)}_{12}&=(1-p)[(1-\tilde\gamma_{12})]^n \\
d^{(n)}_{22}&=-1+(1-p)[\gamma_{22}^n+(1-\gamma_{22})^n]+p[\tilde\gamma_{21}]^n \\
d^{(n)}_{21}&=p [(1-\tilde\gamma_{21})]^n
\end{align}
\end{subequations}

\begin{figure}[h]
\centerline {\includegraphics[clip=true,width=10cm, keepaspectratio]{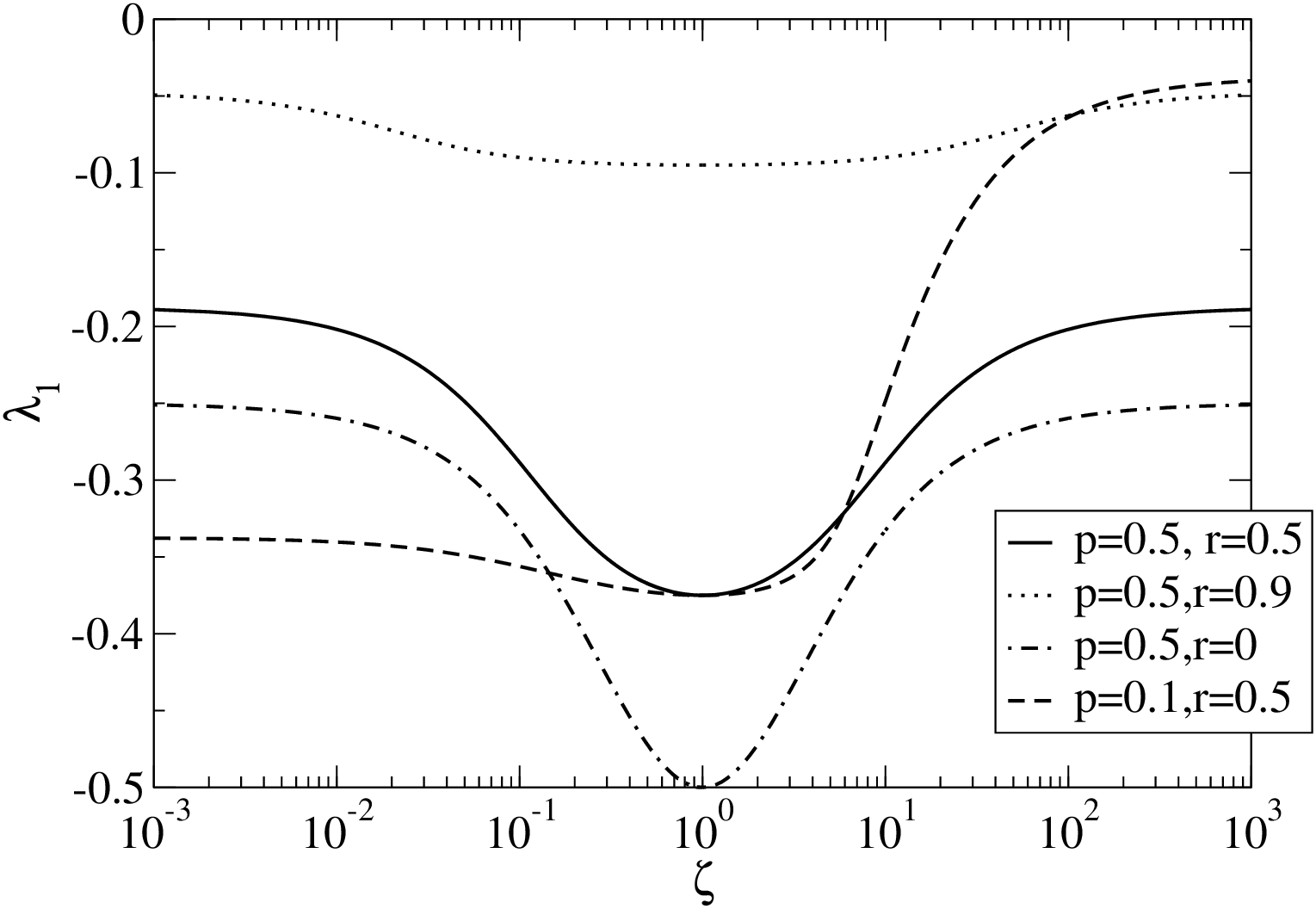}}
\caption{
Maximum eigenvalue $\lambda_1$ of the coefficient matrix of equation
\eqref{secon} for the evolution 
of the second moment, as a
function of the mass ratio $\zeta$. Different cases are shown for
various choices of parameters, $p$ and
$r_{11}=r_{22}=r_{12}=r$.}
\label{fig_autovalore}
\end{figure}

The solution of the system \eqref{secon} is a linear
combination of real exponentials which can be expressed as:

\begin{equation}
\mu^{\alpha}_{2}(t)=A_{\alpha}e^{\lambda_1 t}+B_{\alpha}e^{\lambda_2 t}
\label{mom}
\end{equation}

where $\lambda_1$ and $\lambda_2$ are respectively the less negative
and the more negative eigenvalue of the secular equation associated
with the linear system \eqref{secon}. Their computation
is rather tedious and due to the presence of many control parameters
not particularly illuminating, a part from simple limiting cases, for
which one can extract useful information. For this reason in most
cases we shall give their values by solving numerically the associated
secular equation. An example of the dependence of the larger
eigenvalue $\lambda_1$ from the control parameters is shown in
Fig. \ref{fig_autovalore}.
 
For vanishing first moments the global granular temperature can be
defined as:

\begin{equation}
T_g=p T_1+(1-p) T_2
\label{global}
\end{equation}
 
where $T_{\alpha}=m_{\alpha}\mu^{(2)}_{\alpha}/2$ represents the
partial granular temperature of species $\alpha$.

Notice that in the limit of an elastic system (all
$\gamma_{\alpha\beta}=0$ but arbitrary $\zeta$ and $p$) the eigenvalue
$\lambda_1=0$ reflects the energy conservation. On the other hand, for
non vanishing inelasticity, i.e. values of
$\frac{1}{2}\geq\gamma_{\alpha\beta}>0$, the energy is dissipated at a
finite rate. Consider, for instance, an arbitrary composition $p$, but
$\gamma_{\alpha\beta}=\gamma$ and $\zeta=1$, a situation which
describes the approach to the scaling regime of two subsystems $1$ and
$2$ initially prepared at two different initial granular
temperatures. One finds that the ``energy'' eigenvalue is $\lambda_1=2
\gamma(\gamma-1)$ while $\lambda_2=-(1-\gamma^2)$. To appreciate the
role of $\lambda_2$ let us consider the ratio of the granular
temperatures:

\begin{equation}
\frac{T_1}{T_2}=\zeta  \frac{A_1}{A_2}\frac{1+\frac{B_1}{A_1} e^{(\lambda_2-
\lambda_1) t }}
{1+\frac{B_2}{A_2} e^{(\lambda_2-
\lambda_1) t }}
\label{ratio}
\end{equation}

Relation \eqref{ratio} means that the resulting ``thermalization''
time, viz. the time spent to reach the homogeneous cooling regime is
proportional to $\tau=(\lambda_2-\lambda_1)^{-1}=(1-\gamma)^{-2}$,
i.e. is related to the difference between the two eigenvalues and
attains its maximum for the largest inelasticity $(\gamma=1/2)$. The
asymptotic value of the temperature ratio is given by:

\begin{equation}
\frac{T_1}{T_2}=
\zeta \frac{(d^{(2)}_{11}-d^{(2)}_{22})+\sqrt{(d^{(2)}_{11}-d^{(2)}_{22})^2+
4 d^{(2)}_{12} d^{(2)}_{21}}}{2 d^{(2)}_{21}}
\label{rapporto}
\end{equation}

\begin{figure}[h]
\centerline {\includegraphics[clip=true,width=10cm, keepaspectratio]{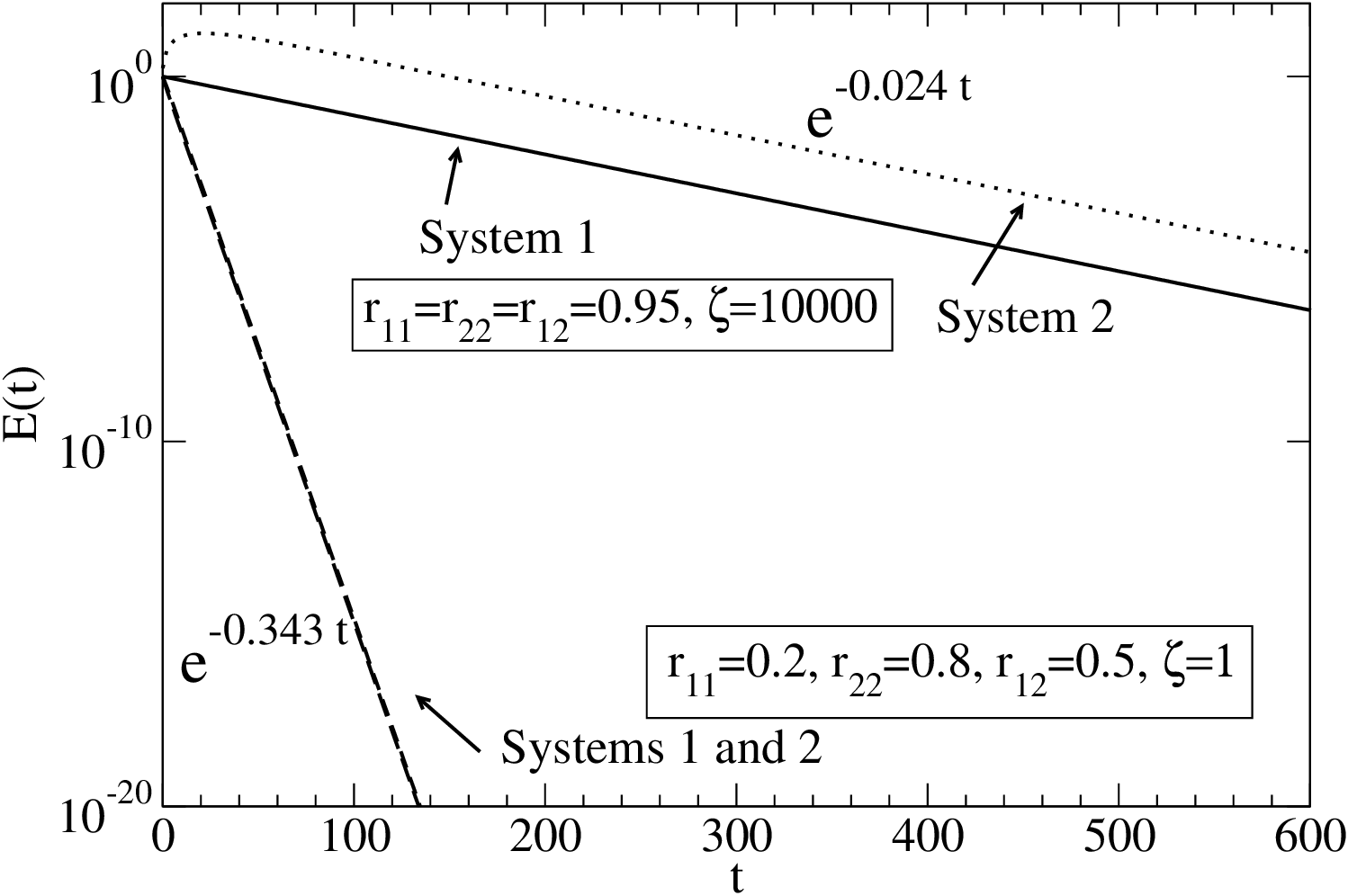}}
\caption{Plot of the average kinetic energy per particle versus collision time.
The two upper curves display the energies of system 1 and respectively. Notice
the initial growth of the average energy of the light species 2. The lowest
curve represent a system, whose temperature ratio is 1}
\label{fig_energy}
\end{figure}

It is clear from eq.\eqref{rapporto} that the two granular
temperatures of the homogeneous cooling system are in general
different (see fig. \ref{fig_energy}) as already 
observed in a different model in
ref. \cite{Dufty}.

In the case of an equimolar mixture ($p=1/2$) with equal inelasticity
parameters ($\gamma_{\alpha\beta}=\gamma$), but arbitrary mass ratio
the temperature ratio is particularly simple:

\begin{equation}
\frac{T_1}{T_2}=
\frac{(\zeta^2-1)\gamma+\sqrt{(1-\zeta^2)^2\gamma^2+4
(1-\gamma)^2\zeta^2}}{2(1-\gamma)\zeta}
\label{mass}
\end{equation}

\begin{figure}[h]
\centerline {\includegraphics[clip=true,width=10cm, keepaspectratio]{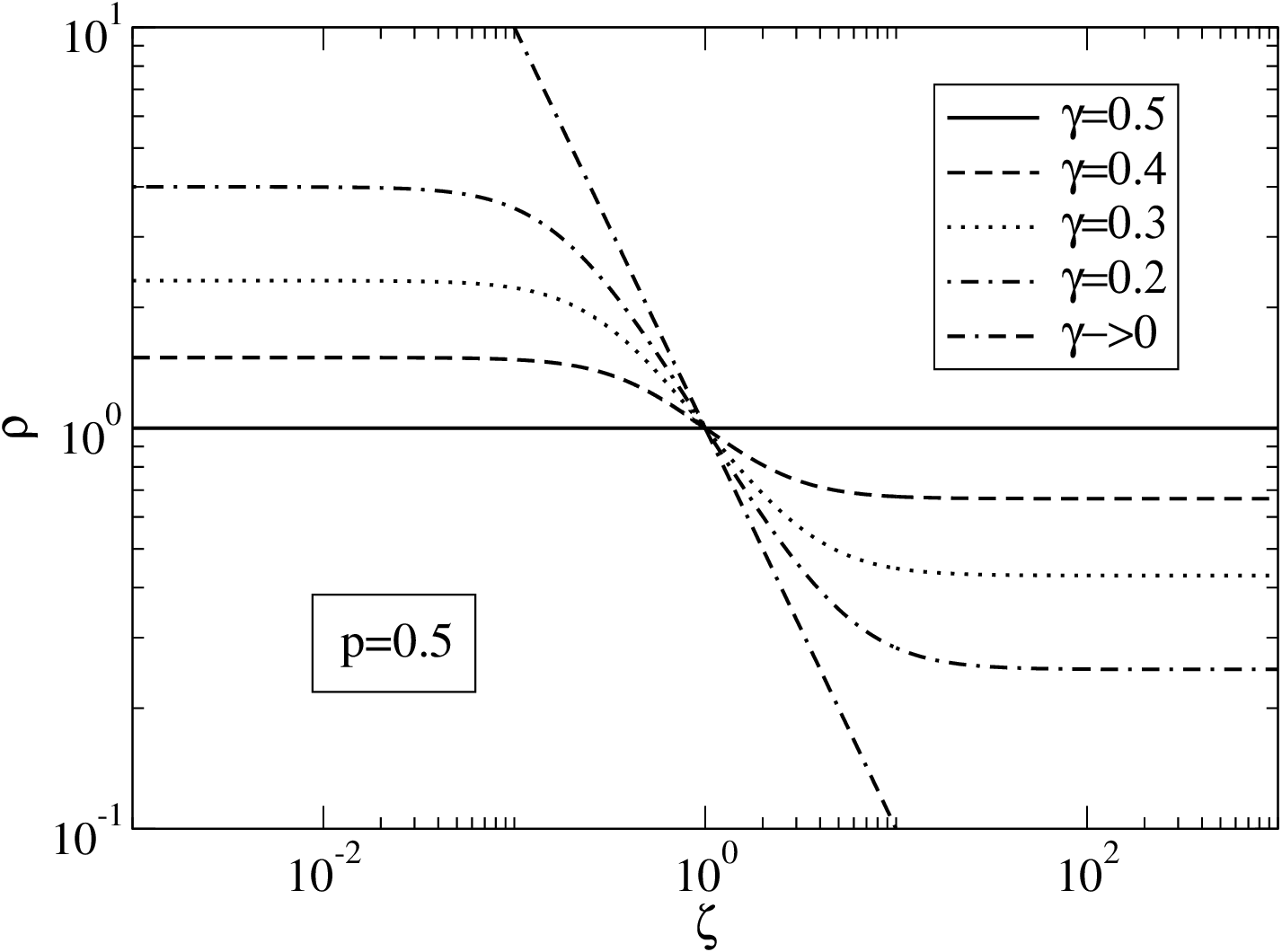}}
\caption{
Asymptotic
ratio $\rho=\mu_2^{(1)}/\mu_2^{(2)}$ of second
moments, as a function of the
mass ratio $\zeta$, for different values of the inelasticity parameter
$\gamma_{11}=\gamma_{22}=\gamma_{12}=\gamma=(1-r)/2$, where $r$ is the
restitution coefficient. $p=0.5$ for all the curves. The perfectly 
inelastic case $\gamma=1/2$ corresponds to equal second moments
for every value of the mass ratio.}
\label{fig_rho}
\end{figure}

\begin{figure}[h]
\centerline {\includegraphics[clip=true,width=10cm, keepaspectratio]{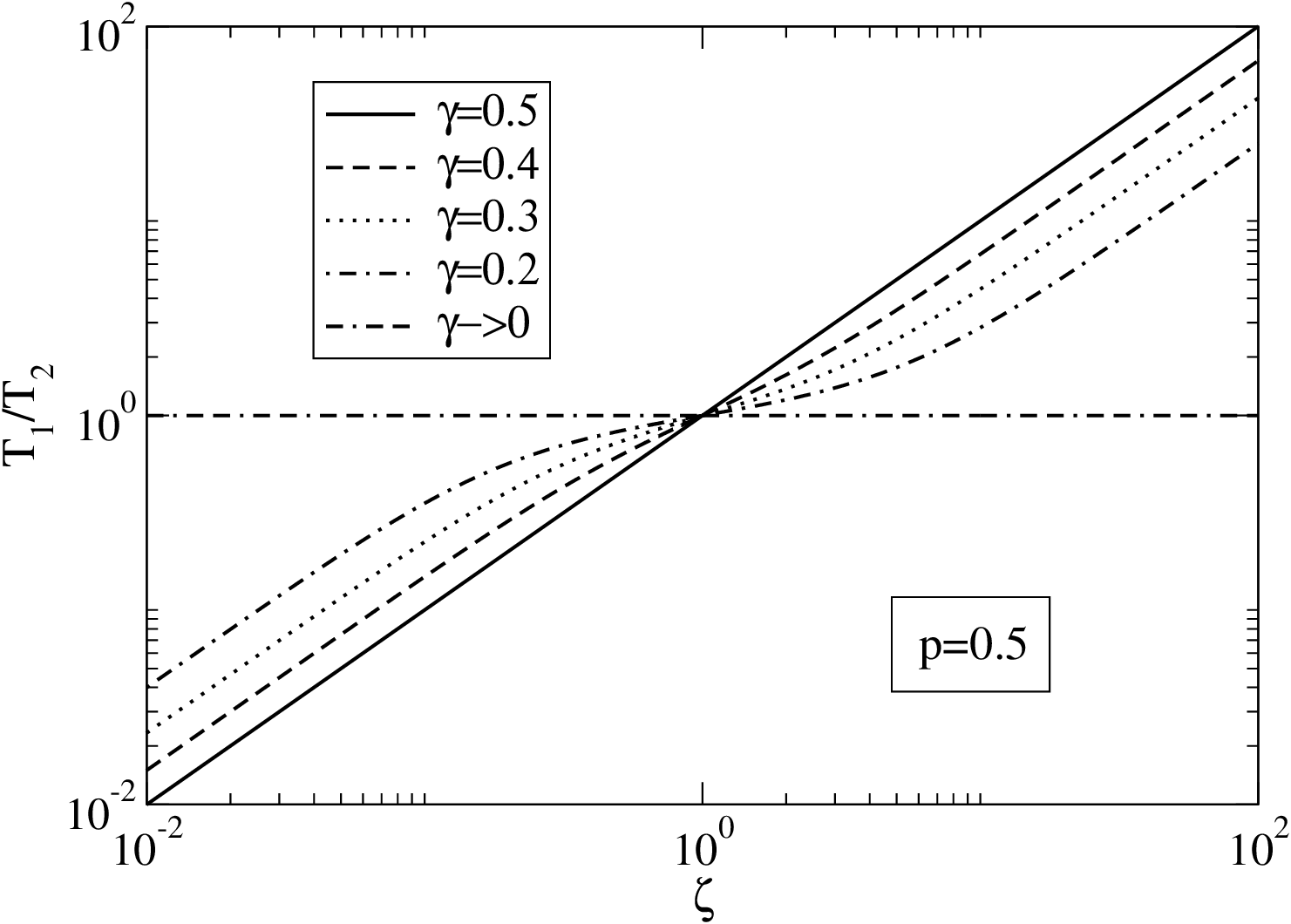}}
\caption{
Ratio $T_1(\infty)/T_2(\infty)$ between asymptotic temperatures of the
two components of the mixture, as a function of the mass ratio
$\zeta$, for different values of the inelasticity parameter
$\gamma_{11}=\gamma_{22}=\gamma_{12}=\gamma=(1-r)/2$, where $r$ is the
restitution coefficient. $p=0.5$ for all the curves. The perfectly
elastic case is the horizontal line, which represents 
energy equipartition.}
\label{fig_t1t2}
\end{figure}

Notice that the second moment ratio $\rho=\frac{\mu^1_2}{\mu^2_2}
=\frac{T_1}{ \zeta T_2}$ approaches $\frac{\gamma}{1-\gamma}$ for
$\zeta \to \infty$ and $\frac{1-\gamma}{\gamma}$ for $\zeta \to
0$. For small departures from the value $\zeta=1$ a good approximation
is:

\begin{equation}
\rho \simeq 1- \frac{1-2 \gamma}{1-\gamma}(\zeta-1)
\end{equation}

i.e. the temperature ratio increases linearly as a function of $\zeta$
and deviates from the equipartition value $1$ of an ideal gas mixture
for non vanishing values of the inelasticity parameter $\gamma$.  In
Figs. \ref{fig_rho} and \ref{fig_t1t2} we display the moment ratio
and the temperature ratio respectively for different values of the
mass ratio and of composition.

When one carries on the evaluation of higher order moments one
discovers the following phenomenon, which is the symptom that the
moment expansion is ill defined: the rescaled moments beyond that of
order $m$ diverge, that is to say the decay of the $m$-th moment is
slower than that of the $m/2$-th power of the second moment. Ben-Naim
and Krapivski \cite{BenNaim} found that in a pure system such a
phenomenon occurs for all moments beyond the fourth.  In other words
the kurtosis of the velocity distribution diverges.  This is the
fingerprint of the presence of an inverse power-law tail in the
velocity distribution function.  The situation became clearer after
the discovery of a scaling solution
\cite{Balda2} as we shall discuss below. 

A simple check shows that the state of affairs remains the same in the
case of a multicomponent mixture. To show that, we write the evolution
equations for the fourth moments:
 
\begin{subequations} 
\label{qua}
\begin{align}
\partial_t \mu^{(1)}_{4}&=d^{(4)}_{11} \mu^{(1)}_{4}+ d^{(4)}_{12} \mu^{(2)}_{4}
+a_{11}(\mu^{(1)}_{2})^2+a_{12} \mu^{(1)}_{2}\mu^{(2)}_{2} \\
\partial_t \mu^{(2)}_{4}&=d^{(4)}_{21} \mu^{(1)}_{4}+ d^{(4)}_{22} \mu^{(2)}_{4}
+a_{22}(\mu^{(2)}_{2})^2+a_{21} \mu^{(1)}_{2}\mu^{(2)}_{2}
\end{align}
\end{subequations}

where the coefficients $a_{\alpha \beta}$ are given by:

\begin{subequations}
\begin{align}
a_{11}&=6 p [\gamma_{11}(1-\gamma_{11})]^2 \\ 
a_{22}&=6 (1-p) [\gamma_{22}(1-\gamma_{22})]^2 \\
a_{12}&=6(1-p) [(1-\tilde\gamma_{12})]^2 [\tilde\gamma_{12})]^2 \\
a_{21}&=6 p [(1-\tilde\gamma_{21})]^2 [\tilde\gamma_{21})]^2
\end{align}
\end{subequations}

We have explored the hyper-surface $\zeta=1$ of the five dimensional
parameter space $\gamma_{\alpha\beta},p,\zeta$ by randomly generating
the values of the coupling constants $\gamma_{\alpha\beta}$ and of the
composition $p$ and computed the eigenvalues, $\Lambda_1$ and
$\Lambda_2$, with $\Lambda_1 >\Lambda_2$, of the secular
equation for the fourth moments, observing that the less negative
eigenvalue $\Lambda_1$ is larger than twice the eigenvalue
$\lambda_1$. This is equivalent to say that after a transient the
rescaled fourth moment will begin to diverge as in the one component
case.  Interestingly, for $\zeta \neq 1$ we observed finite rescaled
fourth moments. However, this fact does not imply that for such values
of the parameters the moment expansion of the rescaled distribution
holds to all orders. It only means that an analogous problem appears
at an higher order of the moment expansion, say at the $m$-th
moment. We shall elaborate on such an issue in section V.


\section{The scaling solution}

A clue to understand the divergence of the rescaled higher moments
came recently from the analysis of the pure case. An exact asymptotic
scaling solution of the master equation \eqref{prob} valid in the
asymptotic regime has been found \cite{Balda2}. The merit of such an
explicit solution is to show that the characteristic function does not
possess a moment expansion, because of the presence of a non
analyticity at the origin, $k=0$ as shown by its representation:

\begin{equation}
f(k v_0(t))=e^{-v_0|k|}(1+v_0|k|)
\label{scali}
\end{equation}

where $\lim_{t \to \infty} \hat P(k,t)=f(k v_0(t))$ and $v_0=A
e^{-\gamma(\gamma-1)t}$ is the square root of the second moment.

To appreciate such a result one can return to real velocity space
where the distribution space is of the form:

\begin{equation}
P_s(\frac{v} {v_0(t)})=\frac{2}{\pi}\frac{1}{v_0}\frac{1}{[1+(\frac{v}{v_0})^2]^2}
\label{baldassariana}
\end{equation}

and that the moment beyond the third diverge due to the presence of
the inverse-power $v^{-4}$ tail.  In fact, it is a general principle
that when a singularity of the function in $k$-space approaches the
real axis the behavior of the large $v$ components of its Fourier
transform turns from exponential to power law
\cite{Migdal}\cite{Lighthill}.
 
An immediate check shows that in the two-component case the above
formula fails to provide the correct solution of the master equation,
with the exception of some special cases, which correspond to a ratio
of the second moments equal to $1$.

Let us stress the importance of scaling solutions in the cooling of an
inelastic gas. Their existence means that the distribution is fully
characterized by its second moment and is self similar,
i.e. asymptotically its shape does not change apart from a trivial
rescaling. Moreover, it demonstrates that in the cooling problem the
distribution functions do not look at all like the distribution
functions of an elastic system, which have an infinite number of non
diverging moments. The change induced by the presence of inelasticity
is a singular perturbation, viz. an abrupt qualitative modification of
the statistical properties of the systems.

Do scaling solutions exist in the case of binary mixtures?  To answer
such a question let us assume that for a constant cooling rate there
exists a scaling solution of the form $\hat
P_{\alpha}(k,t)=f_{\alpha}(\mu_2^{(\alpha)}(t) k^2)$. This implies
that the master equation for $\hat P$ can be converted into a
differential equation with respect to the independent variable $k$
with time independent coefficients. In the one component case it can
be cast in the form:

\begin{equation}
-\alpha k \partial_k \hat P(k)+\hat P(k)-\hat P(\gamma k)P((1-\gamma) k)=0
\label{pippo}
\end{equation}

with $\alpha$ a constant. Eq. \eqref{pippo} displays the presence of a
regular singular point at the origin, $k=0$.  Therefore we look for a
particular solution of the the form: $$ y(k)=k^{\sigma} R(k) $$ where
$\sigma$ is the so called indicial exponent and $R(k)$ is a function
which is analytic at $k=0$, whose value can be determined by the
Frobenius method \cite{Bender}, which provides a systematic tool to
compute the solution as a power series of $k$ with non integer
exponents. If $R(k)$ is analytic one can write: $$
y(k)=k^{\sigma}\sum_{n=0}^{\infty}a_n k^n $$

However, instead of employing the general method we found more
convenient to extend to the two component case the abridged version
employed by Ernst and Brito to ascertain the nature of the
high-velocity tails of the distributions
\cite{Ernst2},\cite{Ben3}. Their technique consists in replacing the
small-$k$ Frobenius expansion with its first few terms which contain a
contribution non analytic at the origin and determining
self-consistently the indicial exponent. We assume that the leading
singularities of the Fourier transforms are of the form:

\begin{equation}
\hat P_{\alpha}(k,t)=1-\frac{1}{2} \mu_2^{(\alpha)}(t) k^2+ 
(\mu_2^{(\alpha)}(t)S_{\alpha}|k|^2)^{\sigma/2} + 
(\mu_2^{(\alpha)}(t))^2 Q_{\alpha} k^4
\label{singol}
\end{equation}

with exponent $\sigma$ and amplitudes $Q_{\alpha}$ , $S_{\alpha}$ to
be determined by substituting the above approximation into the master
equation.  The reason for keeping the fourth term will become clear
below.  Notice that the first two terms in eq. \eqref{singol} are
fixed respectively by the normalization condition and by the variance
of the distribution.  In fact, up to the second order
eq. \eqref{singol} is analytic and identical to the Taylor expansion
\eqref{Taylor}.

Substituting eq..\eqref{singol}) into eqs. \eqref{prob}
and equating the like powers of $k$ we obtain the following set of
coupled equations:

\begin{subequations}
\label{s}
\begin{gather}
\begin{split}
S_1&[\partial_t +1-p(\gamma_{11}^{\sigma}+(1-\gamma_{11})^{\sigma})-
(1-p)|\tilde\gamma_{12}|^{\sigma}]  (\mu^{(1)}_2)^{\sigma/2} \\
-S_2&[(1-p)|1- \tilde\gamma_{12}|^{}] (\mu^{(2)}_2)^{\sigma/2}=0
\end{split} \\
\begin{split}
S_2&[\partial_t +1-(1-p)(\gamma_{22}^{\sigma}+(1-\gamma_{22})^{\sigma})-
p|\tilde\gamma_{21}|^{\sigma}] (\mu^{(2)}_2)^{\sigma/2} \\
-S_1&[p|1-\tilde\gamma_{21}|^{\sigma}] (\mu^{(1)}_2)^{\sigma/2}=0
\end{split}
\end{gather}
\end{subequations} 

To proceed further, we notice that for times much longer than the
``thermalization'' time we can safely approximate the second moments
with their projection along the eigenvector corresponding to the
larger eigenvalue. Doing so the equations \eqref{s} reduce
to a homogeneous system for the variables $S_{\alpha}$. Only in
correspondence of special values of the exponent $\sigma$ the
determinant of the coefficients is zero and there exist non vanishing
solutions. In other words we are requiring a solvability
condition. The resulting indicial equation is highly non linear in the
unknown $\sigma$, but its numerical solution is straightforward. It
reads:

\begin{multline}
\label{solvab}
[\frac{\sigma}{2} \lambda_1+1-p[\gamma_{11}^{\sigma}+
(1-\gamma_{11})^{\sigma}]-(1-p)
|\tilde\gamma_{12}|^{\sigma}] \\
\times [\frac{\sigma}{2} \lambda_1+1-(1-p)[\gamma_{22}^{\sigma}+
(1-\gamma_{22})^{\sigma}] \\
-p|\tilde\gamma_{21}|^{\sigma}]
-p(1-p)|1-\tilde\gamma_{12}|^{a}|1-\tilde\gamma_{21}|^{a}=0
\end{multline}

We observe that eq. \eqref{solvab} has always two solutions $\sigma=0$
and $\sigma=2$ for any choice of the parameters
$\gamma_{\alpha,\beta}$, $\zeta$ and $p$. In fact, these two values
correspond to the zeroth and second moment. The non trivial value
$\sigma>2$ represents the indicial exponent associated with the
singularity. Somehow surprisingly, we find that such a value of
$\sigma$ is distributed continuously in the interval $(2<\sigma<3)$ as
a function of the control parameters. In other words it means that the
inverse-power law tails of the distribution are sensitive to the
composition and to the nature of the interactions in the mixture. We
have not found any simple dependence of the exponent $\sigma$, on the
control parameters. Nevertheless, for small asymmetry we observe that
$\sigma$ seems to deviate from the exponent $\sigma=3$ of the pure system
quadratically with the temperature ratio.

Knowing just the first three terms of the series \eqref{singol}, apart
from the value of the constant $S_1=S_2$ which is still at our
disposal, we make the hypothesis that they represent the truncation of
the expansion of the following characteristic function:

\begin{equation}
\hat P_{\alpha} (k b_{\alpha})=
\frac{2} {\Gamma(\nu)}(\frac{k b_{\alpha}}{2})^{\nu}
K_{\nu}(k b_{\alpha} )
\label{bessel}
\end{equation}

where $\nu=\sigma/2$ and $K_{\nu}(z)$ is the modified Bessel function
of the second kind of order $\nu$. To render the matter clearer we
employ the the following Frobenius series representation
\cite{Abramowitz}:

\begin{equation}
\label{Frob}
\begin{split}
\frac{2} {\Gamma(\nu)}(\frac{k b_{\alpha}}{2})^{\nu}
K_{\nu}(k b_{\alpha} )&=\frac{2} {\Gamma(\nu)}\frac{\pi}{sin(\pi\nu)}
\sum_{n=0}^{\infty} \frac{(\frac{k b_{\alpha}}{2})^{2 n}}{n!}
[\frac{1} {\Gamma(n+1-\nu)}- \frac{\frac{(k b_{\alpha}}{2})^{2 \nu}} 
{\Gamma(n+1+\nu)}]  
\end{split}
\end{equation}
considering just the first few terms
\begin{equation}
\hat P_{\alpha} (k b_{\alpha})
\simeq 1- \frac{1} {(\nu-1)}  
(\frac{k b_{\alpha}}{2})^2 - (\frac{k b_{\alpha}}{2})^{2\nu}
\frac{\Gamma(1-\nu) } {\Gamma(1+\nu)}+ (\frac{k b_{\alpha}}{2})^{4}
\frac{\Gamma(1-\nu) } {2 \Gamma(3-\nu)} 
\label{approxi}
\end{equation}

Thus one can see that the first terms of the series \eqref{singol} and
\eqref{Frob} have the same exponents. Therefore by suitably choosing
the coefficients of \eqref{singol} we can obtain the Bessel function.
In reality, we do so because we are led by the solution of the pure
case which corresponds to $\nu=3/2$. Inserting such a value and
employing the following asymptotic expansion (for $z >0)$ $$
K_{3/2}(z)=\sqrt{\frac{\pi}{2 z}} e^{-z}(1+z^{-1}) $$ after
substitution in eq. \eqref{bessel} we obtain the solution
\eqref{scali}.

If we insist that the identification between the series \eqref{singol}
and the modified Bessel function holds for arbitrary $\nu$ we obtain
by Fourier transforming eq. \eqref{Frob}  the
following approximation to the distribution function
\cite{Gradshtein}:

\begin{equation}
P_{\alpha} (\frac{v}{v_{\alpha}(t)})=
\int_{-\infty}^{\infty} 
\frac{dk}{2 \pi} e^{-ikv} \hat P_{\alpha} (k b_{\alpha})=
\int_{-\infty}^{\infty} 
\frac{dk}{2 \pi} e^{-ikv}
\frac{2} {\Gamma(\nu)}(\frac{k b_{\alpha}}{2})^{\nu}
K_{\nu}(k b_{\alpha} ) 
\label{scali2}
\end{equation}

i.e.:

\begin{equation}
P_{\alpha} (\frac{v}{v_{\alpha}(t)})=
\frac{1}{b_{\alpha}\sqrt{\pi}}
\frac{\Gamma(\nu+\frac{1}{2})}{\Gamma(\nu) } 
\frac{1}{[1+(\frac{v}{b^2_{\alpha}})^2]^{\nu+\frac{1}{2}}}
\label{scali1}
\end{equation}

where $v^2_{\alpha}=\frac{b^2_{\alpha}}{2(\nu-1)}$ represents the
second moment.

\begin{figure}[h]
\centerline {\includegraphics[clip=true,width=10cm, keepaspectratio]{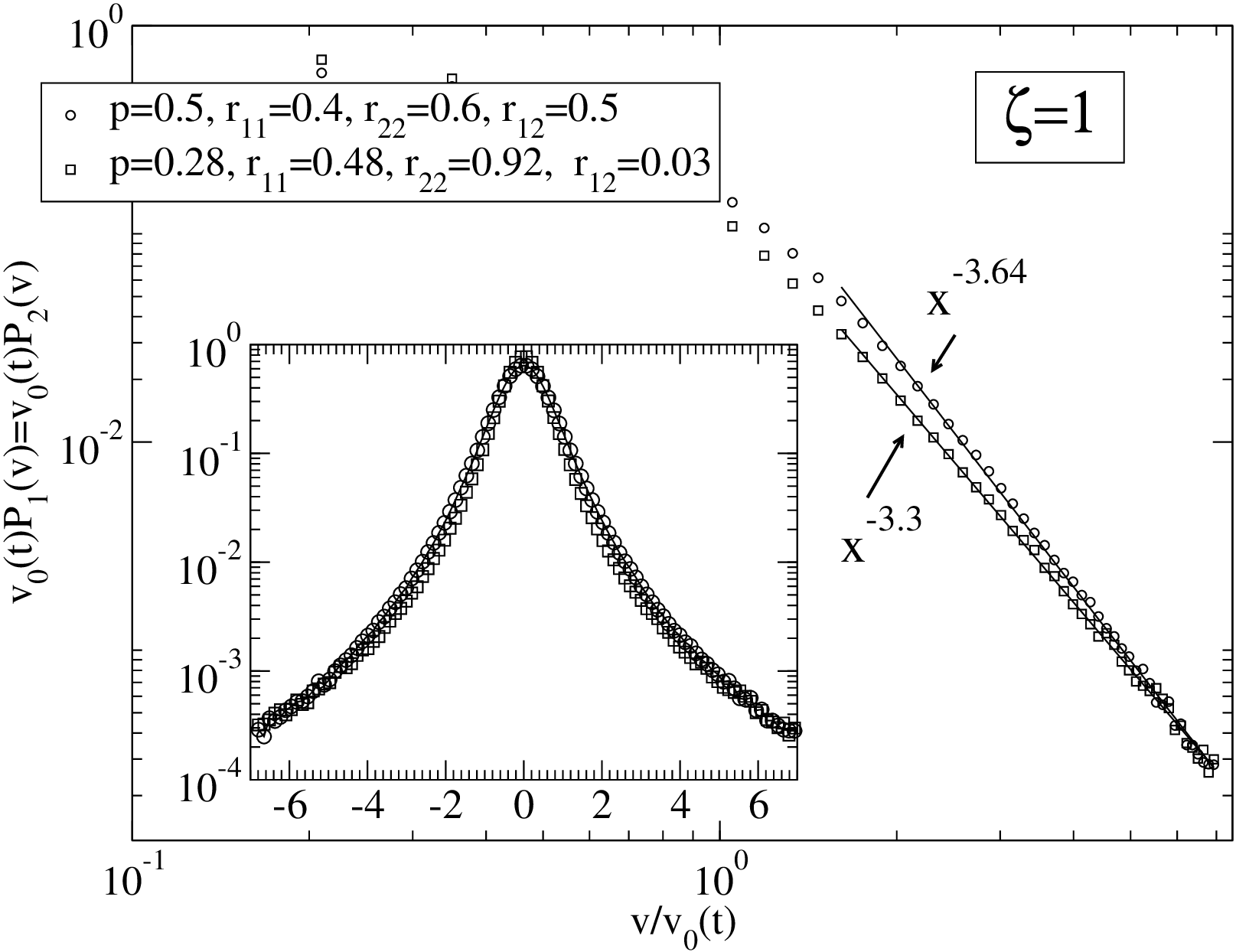}}
\caption{
Rescaled asymptotic velocity distributions from numerical simulations
of the model with mass ratio $\zeta=1$ and different values of the
restitution coefficients. In the inset the whole distributions are
shown, in the main graph the tails are magnified in log-log scale,
and fitted with an inverse power law. The predictions obtained
by solving the indicial equation are $\sigma=-3.86$ for the upper curve
and $\sigma=-3.21$ for the lower curve.}
\label{fig_v_zeta1}
\end{figure}

Such an approximate form clearly displays the inverse power law tails
with the characteristic exponent $\sigma+1=2 \nu +1$.  For a pure
system since $\sigma \to 3$ it reduces to the known solution
\cite{Balda2}. Numerically we found that such an approximation gives
excellent results with an error which is compatible with the
uncertainty of the numerical data as shown in
Fig.\ref{fig_v_zeta1}.  Notice that the indicial equation does
not give the value of $S_{\alpha}$, but this has been fixed by our
ansatz \eqref{bessel} . On the other hand we might ask how good is the
ansatz.  To clarify such an issue we substitute the expansion
\eqref{singol} into the scaling form of the master equation and find
two linear inhomogeneous algebraic equations for the two unknown $Q_1$
and $Q_2$. Without giving all the details we notice that these
equations have certainly non zero solutions in virtue of the fact
observed above, eqs. \eqref{qua} that the eigenvalues of the
fourth moment is not twice the eigenvalues of the of the second
moment.  The numerical solution of the linear system gives two values
of the $Q$'s which in general are different ($Q_1 \neq Q_2$) for non
special values of the parameters, hence the ansatz
represents only an approximation. The consequences of this small
discrepancy are probably of minor importance for the cases considered
in the present section.  The deviation of the true series from the
series representation of the approximate solution, might manifest
itself in a small asymmetry in the two rescaled distribution
functions, and this might involve the region of small values of $v$.
In principle there exist the possibility of constructing a better
solution via the Frobenius method. However, we believe that one could
hardly find a closed form of the distribution functions via an inverse
Fourier transform of simplicity comparable to that of eq
\eqref{scali}. Finally, let us remark that the fairly good agreement
between our approximation and the numerically computed solution stems
from the fact that the form we propose embodies not only the three
following basic ingredients: i) the normalization condition, ii) the
correct value of the variance and iii) the appropriate tails of the
distributions, but also some properties of the exact distribution
which one can guess on the basis of pure physical reasoning.  These
properties are that $P_{\alpha}(v)$ is symmetric w.r.t. $v$, is
monotonically decreasing for $v>0$, and is smooth.  Such assumptions
hugely restrict the class of all possible candidate distributions
compatible with the first few terms of the expansion. In practice we
match the small-$k$ behavior with the small $r$ behavior, which is
equivalent to the large $k$-behavior.


\section{Higher order singularities}

So far we have employed the approximate expansion \eqref{singol} and
found that on the hypersurface $\zeta=1$ the indicial exponent
$\sigma$ lies in the continuous interval $2<\sigma\leq 3$
Correspondingly the distributions possess finite second moments, but
diverging fourth moments. On the other hand for $\zeta \neq 1$ as
anticipated in section III it is possible to observe different kinds
of behaviors.  The feature which makes the difference is what might be
called the ``isotopic effect'', i.e. induced solely by the difference
of masses in a binary mixture of particles, otherwise identical. One
observes the following phenomenon: all rescaled moments
$\mu^{(\alpha)}_s/(\mu^{(\alpha)}_2)^{s/2}$ up to the $m$-th order are
asymptotically finite, but diverge beyond $m$.

This is indicative of solutions characterized by very large values of
$\sigma$. In other words, the allowed interval of values of $\sigma$
is expanded. This physically means that the exponent $\sigma$ evolves
from a pronounced inverse power law to a Gaussian-like behavior as $\zeta$
deviates from unity.  Of course, the crossover from one regime to the
other is determined by the inelasticity and by the mass ratio in the
first place, but the value of the exponent does not depend on any
simple way from the control parameters.

\begin{figure}[h]
\centerline {\includegraphics[clip=true,width=10cm, keepaspectratio]{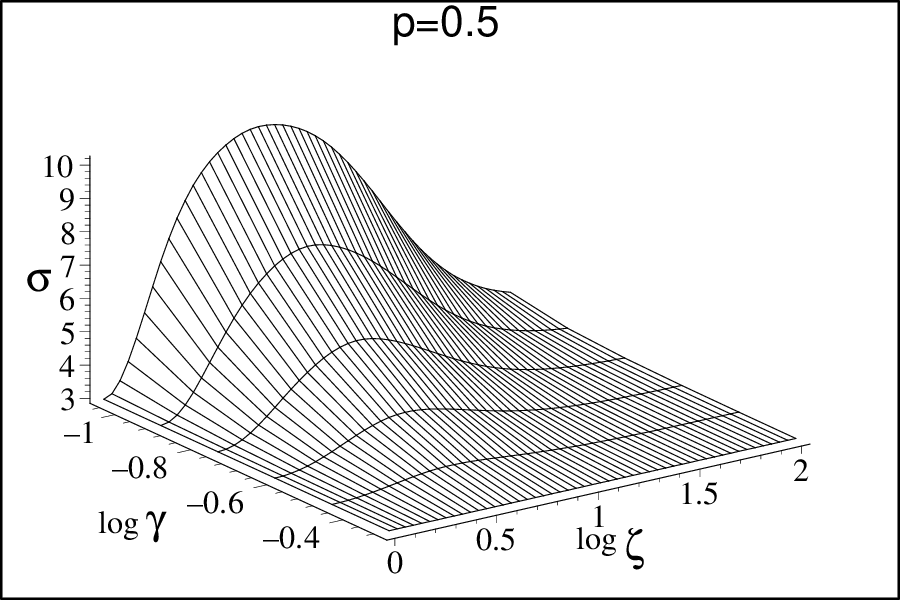}}
\caption{
Power of the singularity $\sigma$ of the solution of the coupled master
equations for the model, as a function of the mass ratio $\zeta$ and
the inelasticity parameter
$\gamma_{11}=\gamma_{22}=\gamma_{12}=\gamma=(-r)/2$, with $r$ the
restitution coefficient, with $p=0.5$.}
\label{fig_sigma}
\end{figure}

Numerically we have considered different coupling parameters and
obtained the results shown in Fig.\ref{fig_sigma} showing the trend
that the exponent of the tails increases with decreasing inelasticity
and with the difference $|\zeta-1|$.
 
\begin{figure}[h]
\centerline {\includegraphics[clip=true,width=10cm, keepaspectratio]{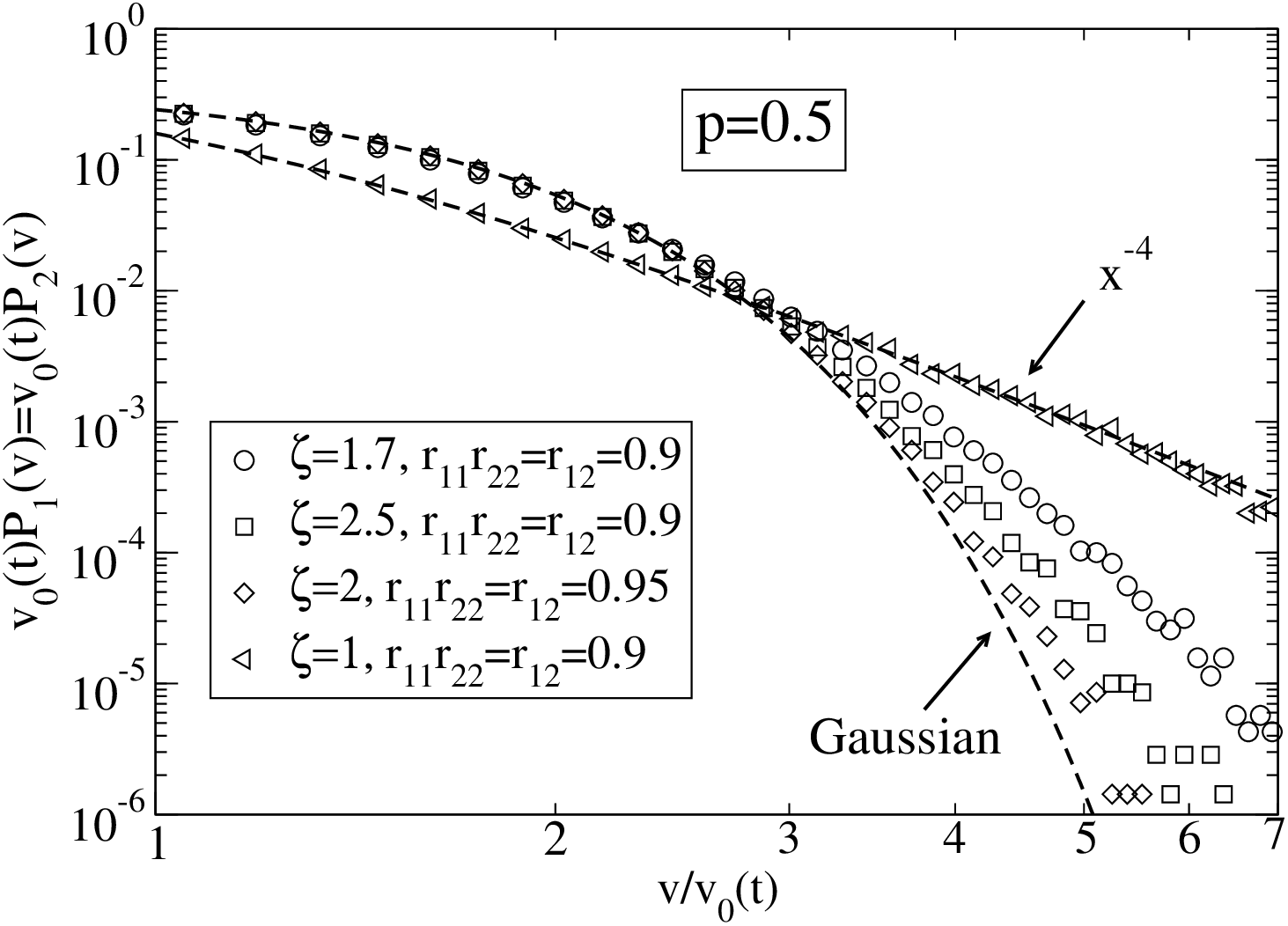}}
\caption{
Tails of the rescaled asymptotic velocity distributions from numerical
simulations of the model with number ratio $p=0.5$ and different
values of the restitution coefficients and of the mass ratio
$\zeta$. The pure case $\zeta=1$ has the exact asymptotic solution
$P(x)=(2/\pi)(1+x^2)^{-2}\sim x^{-4}$. The Gaussian is shown as a guide
for the eye.}
\label{fig_v_zeta_vari}
\end{figure}

In Fig. \ref{fig_v_zeta_vari} we have plotted the distributions
obtained with different values of $\zeta$, showing the change of the
exponent of the tails with $\zeta$. We must stress the difficulty of
obtaining clear numerical results for the exponents of the power law
tails in the case of large values of the singularity $\sigma$ (or of
the absolute value of the exponent itself, which is $\sigma+1$): in
fact high inverse-power law tails need far larger statistics to be
appreciated; moreover, it appears (numerically) that for large
(negative) values of the expected power, the tail appears later in the
$v/v_0$ domain: this means that with a finite statistics one can
measure powers smaller, in absolute value, than those expected from
the analysis of indicial equations. This phenomena has been observed
also in~\cite{Ernst2}.

To simplify the analysis, let us consider first the master equation
for large values of $\zeta$ and equal inelasticity parameters. One
sees that as $\zeta \to \infty$ the first equation for the
distribution $P_1$ is asymptotically decoupled from $P_2$.

\begin{subequations}
\begin{align}
\lim_{\zeta \to \infty} \tilde\gamma_{12}=1 \label{g12} \\
\lim_{\zeta \to \infty} \tilde\gamma_{21}=2 \gamma -1 \label{g21} \\
\end{align}
\end{subequations}

For small inelasticity $\gamma$, the quantity $\tilde\gamma_{21}$ is
negative as if the restitution coefficient for the collision $(2-1)$
were larger than one, while $1-\tilde\gamma_{12}=0$. Under such
conditions, the equation for the distribution function of species $1$
becomes effectively decoupled from species $2$ as it can be
appreciated by substituting \eqref{g12}-\eqref{g21} in eq.
\eqref{fou}:

\begin{equation}
\label{limit}
\begin{split}
\partial_t \hat P_1(k,t)&=-\hat P_1(k,t)+ p \hat P_1(\gamma k,t) 
\hat P_1((1-\gamma)k,t) \\
&+(1-p) \hat P_1(k,t) \hat P_2(0,t) 
\end{split}
\end{equation}

Taking into account the fact that $\hat P_2(0,t)=1$ one recognizes
immediately the equation for the p.d.f. of a pure system, i.e. of the
form given by eq. \eqref{baldassariana}.  The decay rate of the energy
is given by $\lambda_1=2 \gamma(\gamma-1)p$.  What happens to the
light component? Within this limit the evolution equation for $\hat
P_2(k,t)$ looks very different from that of $\hat P_1(k,t)$. In fact,
the numerical solutions indicate the existence of tails with very
large values of $\sigma$.  How can we find this value of the exponent,
knowing that the tails of the species $1$ diverge as $v^{-4}$? The
trick is to substitute in \eqref{fou2} the expansion \eqref{singol}
and neglect the contribution stemming from $\hat P_1(k,t)$ because it
corresponds to another degree of singularity. At the end one obtains
the new indicial equation:

\begin{equation}
2 \gamma(\gamma-1) p \frac{\sigma_2}{2}+1-(1-p)[
\gamma^{\sigma_2}+(1-\gamma)^{\sigma_2}]+
p |1-2 \gamma|^{\sigma_2}=0
\label{sing2}
\end{equation} 

which for small values of $\gamma$ has a solution $\sigma_2 \simeq 
\frac{1}{p \gamma(1-\gamma)}$. That is to say  for 
a quasi-elastic system the exponent of the singularity diverges
and all moments exist. To understand such a result we can reconsider
the rule \eqref{collision} and see that for $\zeta \to \infty$ the
collisions between unlike particles do not change neither the energy
nor the velocity of the heavy species, whereas it changes the velocity of 
particles $2$ in the following way:

\begin{equation}
{v'}^{(2)}_i=v^{(2)}_i+\zeta(v^{(1)}_j-v^{(2)}_i)
\end{equation}

i.e. the species $1$ acts on $2$ as a stochastic noise, since $
v^{(1)}$ is randomly distributed according to $P_1(v,t)$. Such a
phenomenon is peculiar of the inelastic system under scrutiny and
represents a sort of ``reversed Brownian'' motion in which the heavy
particles act as a heat bath for the light particles.

\begin{figure}[h]
\centerline {\includegraphics[clip=true,width=10cm, keepaspectratio]{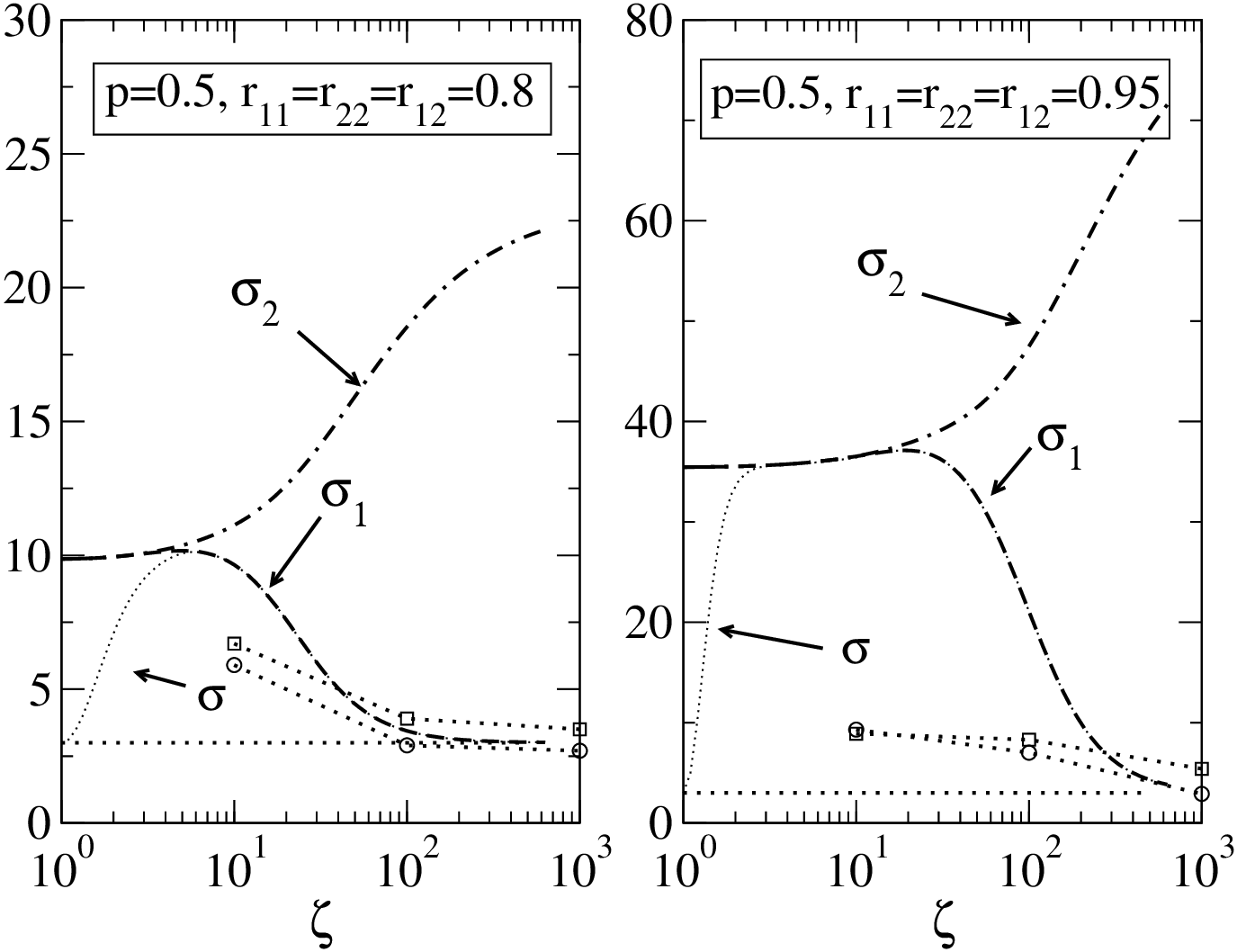}}
\caption{
Indicial exponents $\sigma$, $\sigma_1$ and $\sigma_2$ of the singularity of the
solutions of the coupled master equations \eqref{fou} and of the
singularities of the solutions of the non-coupled master equations
\eqref{limit} and \eqref{sing2}, respectively, as 
functions of the mass ratio $\zeta$. The squares and circles refer to
the exponents obtained from the numerical simulations (to be compared
with $\sigma_1$ and $\sigma_2$ respectively): the discrepancy with the
theoretical predictions is due to the slow convergence, for large
$\sigma$'s, of the tails to the asymptotic value, as discussed in the
text.}
\label{fig_sigma_diff}
\end{figure}

In the case of finite $\zeta$ we have solved the indicial equation
\eqref{solvab} and constructed the curves shown in Fig.
\ref{fig_sigma_diff}. For small values of the inelasticity $\gamma$ the
surface raises steeply as $\zeta$ departs from $1$, attains a maximum
and then decreases again reaching the asymptotic value $\sigma=3$ for
very large values of $\zeta$, as predicted by our asymptotic analysis
(see fig. \ref{fig_sigma}).
For larger values of the inelasticity such isotopic effect is less
pronounced. This reentrant behavior of $\sigma$ with $\zeta$ is
mirrored by an analogous behavior of the rescaled moments.

To conclude, we remark that for large values of $\nu$ the coefficients
of the series expansion in powers of $v$ of order $n<\nu$ are very
close to the corresponding coefficients of the series expansion of the
Gaussian, and only for $n>\nu$ the power law behavior of the tails
becomes manifest. This means that numerically it might be very
difficult to detect such a region.  Finally for large $\nu$, which
correspond to elastic systems, one recovers the Gaussian distribution:

\begin{equation}
\lim_{\nu \to \infty}
\frac{1}{b_{\alpha}\sqrt{\pi}}
\frac{\Gamma(\nu+\frac{1}{2}) }{\Gamma(\nu) } 
\frac{1}{[1+(\frac{v}{b^2_{\alpha}})^2]^{\nu+\frac{1}{2}}}
= \frac{1}{\sqrt{2 \pi}}\frac{1}{v_0} e^{-\frac{v^2}{2 v_0^2}}
\label{scali3}
\end{equation}

The re-entrance phenomenon of the exponent $\sigma$ deserves a
comment. In fact we have seen that $\sigma$ controls the tails of the
distribution of the heavy component only. In fact, even for $\zeta$
finite, but larger than some value $\zeta_{cr}$ one observes a
bifurcation of the exponents. For $\zeta \geq \zeta_{cr}$ the exponent
of the heavy component begin to be different from the exponent of the
light component. Therefore, our approximation consisting in assuming
the same indicial value for both subsystems becomes
untenable. Nevertheless, we can find the correct values of the
exponents $\sigma_1$ and $\sigma_2$, by considering that there is no
interference between the two singularities associated with
$k^{\sigma_1}$ and $k^{\sigma_2}$ and therefore we get two decoupled
indicial equations.  The resulting scenario is quite intriguing. After
the bifurcation we have two separate trajectories obtained by drawing
the values of $\sigma_1$ and $\sigma_2$ against
$\zeta$. Correspondingly the probability distribution functions of the
two subsystems belong to two different critical hyper-surfaces.  In the
limit of $\gamma \to 0$ one subsystem flows to the Gaussian elastic
fixed point, the other flows to the inelastic fixed point.


\section{Relation to Non Extensive Statistical Mechanics}

We have discussed the cooling behavior of a simple model of granular
material and found that the process gives rise to scaling forms of the
probability distribution function. Somehow surprisingly, the
distribution function has an inverse power law decay for large
velocities of the form $v^{-\sigma-1}$ with an exponent $\sigma$ which
depends in a non trivial fashion on the various control parameters and
can vary in the interval $2<\sigma<\infty$ in a continuous way. It is
an intriguing fact that such inverse-power law like behavior looks
undeniably similar to that emerging in the context of the so called
Nonextensive Statistical Mechanics \cite{Tsallis}. To the best of our
knowledge, no other model of comparable simplicity to the present
yields similar laws \cite{Beck} as a result of a dynamical process. We
stress again that none of our results has been derived from
assumptions about the functional form of a Generalized Entropy and an
associated maximum Entropy Principle. Our results follow from a
completely independent approach, namely the exact treatment of
inelastic collisions based on the master equation.  A possible
explanation of why the Nonextensive Statistical Mechanics and our
formulas are so similar can be found in the non extensive properties
of the energy which is a key assumption of NESM and a characteristic
of our granular model. In addition, the mean field type of couplings
of the present model might be a crucial ingredient to determine the
observed inverse power law tails.

To make an example let us take two different systems at the same
temperatures and observe that they do not equilibrate at a temperature
ratio equal to one.  Repeating the experiment with a third system we
would find a different temperature ratio for each new pair we can
form.  This reflects the non conservation of the kinetic energy;
moreover the amount by which it is non conserved depends on the
microscopic details of the two systems we bring together. Classically
the temperature, besides being the average of the kinetic energy, is
also the intensive thermodynamic field conjugated to the entropy and
the latter in turn is related to the distribution function.  This is
not the case of the granular temperature, which does not possess a
definition via the entropic route, but only via the kinetic route. In
NESM, instead, one defines a generalized entropy,

$$
S_q=\frac{1}{q-1}(1-\sum_i p_i^q)
$$

where the $p_i$ are the probabilities associated with the microstates
$i$ of the system. Therefore $S_q$ does not have a unique expression
as in Classical Physics, but varies with the exponent $q$, connected
to $\sigma$ by the relation:

$$
q=1+\frac{2}{\sigma+1}
$$

Thus $S_q$ would be a function of the various couplings of the
particular system under scrutiny.

Summarizing, although the NESM is capable to yield our results, it does
not provide the value of the exponent $q$ so that one needs an
independent procedure to obtain it and compute the appropriate
q-entropy. This non universality of the $q$ exponent means that one
would need a q-entropy for each particular mixture.  Finally, a
challenging situation for which we do not know the answer within the
NESM, is the one inspired by the results of section V, where the
two-components of the mixture are characterized by different power law
tails; even within the same ``experiment'' one has to resort to two
different q-entropies.
  
This state of affairs is just the result of the absence of the Zeroth
Law of Thermodynamics for granular systems, i.e. the fact that one
needs a different thermometer for each particular granular mixture.

Let us anticipate that in the case of a homogeneously heated granular
system our model still predicts a two temperature behavior under
rather general conditions, but with the partial temperatures not
varying in time \cite{inpreparation}. On the other hand, the velocity
distribution function will not show any power-law tails, but will be
very similar to Gaussians and possess all finite moments. Thus, the
power-law behavior is a peculiarity of the cooling process and not a
necessity of non extensivity, at least as far as our model is
concerned.

\section{Conclusions}

We have seen that the Maxwell inelastic mixture displays a continuous spectrum
of exponents characterizing the inverse power-law tails of the
velocity distribution functions.  The previously found solution of the
pure system $p=1$ with tail exponent $\sigma+1=4$ can now be regarded as
special case.  In fact, a whole spectrum of exponents extending from
$3$ to $\infty$ exists.  A natural question to ask is why the
observed exponent $\sigma$ deviates from the value of the pure
system. The naive guess is that, since in the pure case the the
solution \eqref{scali} shows the remarkable feature that its
functional form is independent of the cooling rate, i.e. of the
microscopic inelasticity parameter $\gamma$ and that the tails decay
invariably as $v^{-4}$, the same should be true in the case of a
mixture of particles characterized by different restitution
coefficients. On the contrary, in the mixture case the $\sigma$
exponent varies with the microscopic parameters.  Both these
contrasting aspects, namely a single value of the exponent 
for the pure model and
a continuous spectrum of 
values for the mixture recall the Renormalization Group theory of
Critical Phenomena, where critical systems having different
Hamiltonians ( viz. different $\gamma$'s in our pure case) in the
neighborhood of a fixed point share the same exponents. That is to say
that the evolution drives all systems belonging to the same critical
(hyper)surface towards a common fixed point. This is why all pure
systems exhibit identical asymptotic behaviors in spite of having different
inelasticity parameter $\gamma$
\cite{Goldenfeld}. On the other hand,
mixtures correspond to systems whose trajectory 
lies on a hyper-surface whose points
are attracted towards a fixed point characterized by $\sigma \neq 3$,
because of the presence of other scaling fields.

The model can also be viewed as a quench realized rendering suddenly
inelastic an initially elastic system.  The subsequent cooling process
takes the distribution from its elastic Gaussian fixed point to to the
zero temperature fixed point. The trajectory along which this process
occurs is characterized by peculiar values of the distributions.  The
scaling behavior is associated with the dynamics of the approach to
the fixed point.

Why the scaling solution is selected? Let us make the following
observation: if we consider a mixture with a large number of identical
components, the mixture formalism allows us to describe independently
the evolution of different subsystems. The associated secular equation
for $M$ components will have $M$ eigenvalues, $M-1$ of them
corresponding to the faster modes. The scaling solution corresponds to
the slowest mode and is therefore stable, with respect to
fluctuations.

To conclude, the model we have investigated can be regarded as a kind
of self-organized critical system.  Due to the mean field nature of
the model the fluctuations of a given portion of the system can
influence all other elements, i.e are very long ranged.  In a more
realistic description of the inelastic interaction the scenario
illustrated above will not survive asymptotically, because of the
onset of spatial fluctuations, i.e. of correlations which tend to
erode the high velocity tails. These tails, on the other hand, might
remain observable during the homogeneous cooling regime but will cease
after the formation of spatial gradients in the system
\cite{Balda1}.

Finally, we argue that the two temperature behavior and the existence
of a common cooling rate for the two species and of scaling solutions
is a robust feature even when a more realistic treatment of the
collision term is performed \cite{inprepa2}.

\section{Acknowledgments}

We wish to thank A. Baldassarri, M.Cencini  and A.Vulpiani
for stimulating discussions.

\begin{figure}[h]
\centerline 
{\includegraphics[clip=true,width=10cm, keepaspectratio]{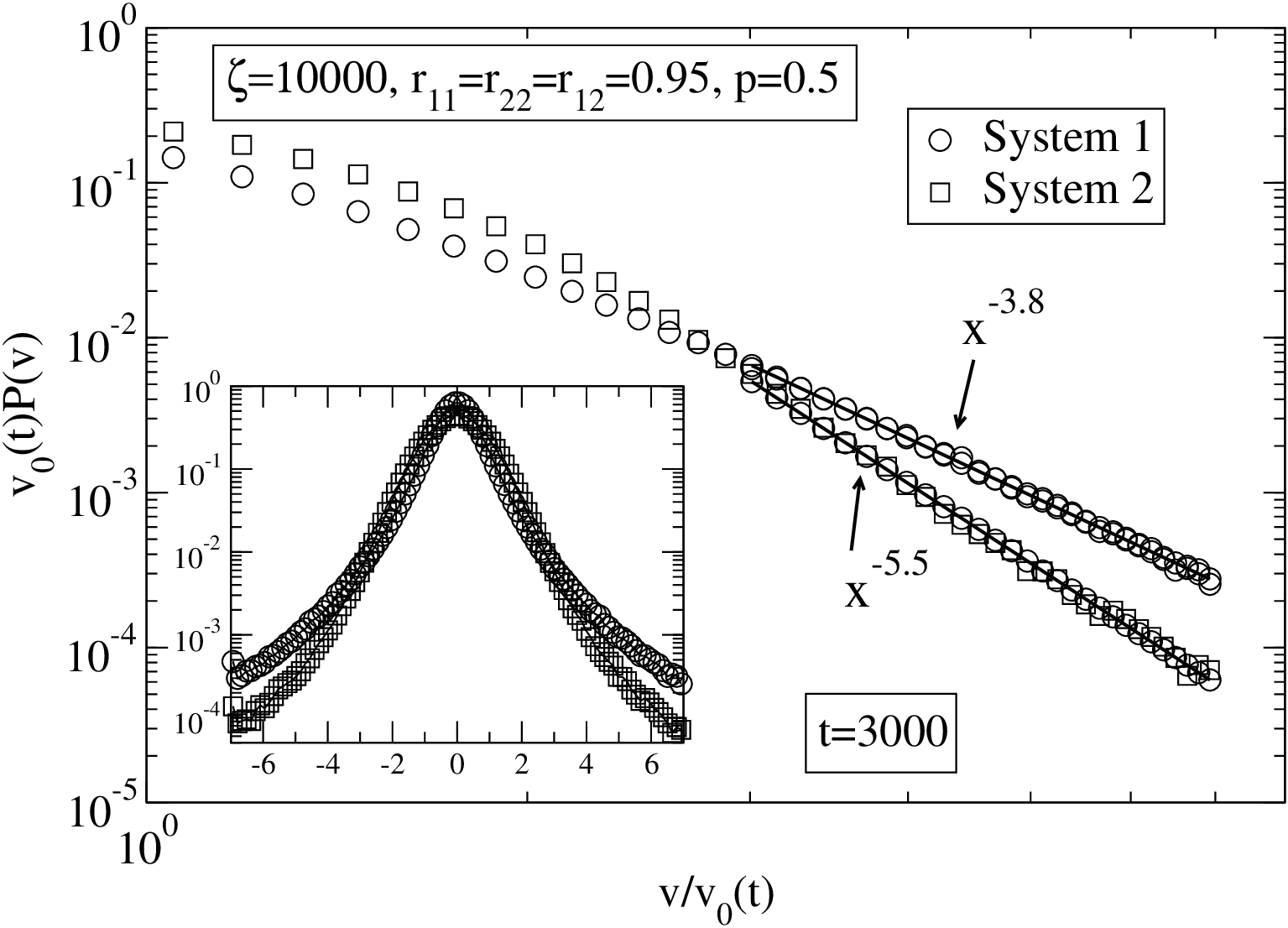}}
\caption{Plot of the rescaled velocity distributions in the case
of a large mass ratio $\zeta=10000$. In the inset the two 
velocity distributions are shown, whereas in the figure the tails are
displayed. The tails are characterized by a different inverse power law.}
\label{fig_v_zeta10000}
\end{figure}

\end{document}